\documentclass[12pt]{article}
\usepackage{latexsym}
\usepackage{amsmath}
\usepackage[dvips]{graphicx}
\usepackage{textcomp}

\newcommand{\be}{\begin{equation}}
\newcommand{\ee}{\end{equation}}
\newcommand{\ba}{\begin{array}}
\newcommand{\ea}{\end{array}}

\author{Fabio Cardone $^{1,2}$, Giovanni Cherubini $^{3,4}$, Roberto Mignani $^{2,5}$, \\ Walter Perconti $^{1}$,
Andrea Petrucci $^{1,5}$, Francesca Rosetto $^{5,6}$ \\ and Guido
Spera $^{7}$ \\ \\ $^{1}$Istituto per lo Studio dei Materiali
Nanostrutturati (ISMN — CNR) \\ Via dei Taurini - 00185 Roma, Italy
\\ $^{2}$GNFM, Istituto Nazionale di Alta Matematica "F.Severi" \\
Citt\`a Universitaria, P.le A.Moro 2 - 00185 Roma, Italy \\
$^{3}$ARPA Radiation Laboratories Via Montezebio - 01100 Viterbo,
Italy \\ $^{4}$ Facolt\`a di Medicina, Universit\`a degli Studi "La
Sapienza" \\ P.le A. Moro, 2 - 00185 Roma, Italy \\
$^{5}$Dipartimento di Fisica "E.Amaldi", Universit\`a degli Studi
"Roma Tre" \\ Via della Vasca Navale, 84 - 00146 Roma, Italy \\
$^{6}$ARPA Chemical Laboratories Via Montezebio - 01100 Viterbo,
Italy \\ $^{7}$CRA - IS.Pa.Ve., Chemical Section \\ Via C.G.
Bertero, 22 - 00156 - Roma, Italy}
\date{}
\title{Neutrons from Piezonuclear Reactions
}
\begin{document}
\maketitle \abstract{We report the results obtained by cavitating
water solutions of iron salts (\emph{Fe(}\emph{Cl)}$_{\emph{3}}$ and
\emph{Fe(}\emph{NO}$_{\emph{3}}$\emph{)}$_{\emph{3}}$) with
different concentrations at different ultrasound powers. In all
cases we detected a neutron radiation well higher than the
background level. The neutron production is perfectly reproducible
and can at some extent be controlled. These evidences for neutron
emission generated by cavitation support some preliminary clues for
the possibility of piezonuclear reactions (namely nuclear reactions
induced by pressure and shock waves) obtained in the last ten years.
We have been able for the first time to state some basic features of
such a neutron emission induced by cavitation, namely: 1) a marked
threshold behavior in power, energy and time; 2) its apparent
occurring without a concomitant production of $\gamma$ radiation.}

\section{Introduction}
Acoustic cavitation of liquids with gas dispersed consists in
subjecting them to elastic waves of suitable power and frequency (in
particular to ultrasounds)~\cite{cg,bren}. The main physical
phenomena occurring in a cavitated liquid (e.g.
sonoluminescence~\cite{met}) can be accounted for in terms of a
hydrodynamic model based on the formation and the collapse of gas
bubbles in the liquid~\cite{cg,bren}. Three different experiments on
cavitation carried out in the last years~\cite{cm1,cm2,cm3} provided
evidence for an anomalous production of intermediate and high mass
number (both stable, unstable and artificial) nuclides within
samples of water subjected to cavitation, induced by ultrasounds
with 20 KHz frequency. Those results together seem to show that
ultrasounds and cavitation are able to generate nuclear phenomena
bringing to modifications of the nuclei involved in the process. A
model able to account for such nuclear reactions induced by high
pressures (called in this paper $piezonuclear~reactions$), based on
the implosive collapse of the bubbles inside the liquid during
cavitation, has been proposed by two of the present authors (F.C.
and R.M.)~\cite{cm4}. Notice that, in the first experiments that we
carried out, proton number was practically conserved, whereas
neutron number was apparently not~\cite{cm1,cm2}. This constitutes
an indirect hint of some sort of neutron production in such
cavitation processes. Since, as is well known, nuclear reactions in
most cases involve neutron emission, it is a fundamental issue to
check whether neutrons are produced indeed in processes possibly
involving piezonuclear reactions. We point out that some experiments
carried out~\cite{tal1}-\cite{tal4} in the last years have shown
that cavitation of deuterated acetone can produce neutrons. In order
to shed some light on this issue of neutron emission during
cavitation, in 2004-2006 we carried out some experiments in which we
cavitated controlled solutions of salts in water at CNR National
Laboratories (Rome 1 Area) and Italian Armed Forces technical
facilities. We focused our attention on ionising radiation and
neutron emission. The details of these experiments are reported in
the following.

\section{Experimental Equipment}

The employed ultrasonic equipment was the robust ultrasound welder
DN20/2000MD by Sonotronic~\cite{sono}. We slightly modified the
piezoelectric and the sonotrode configuration in order to provide
the equipment with a compressed air cooling system which allowed it
to work for 90 minutes without stopping, at a frequency of 20 KHz.
As cavitation chamber, we used a Schott Duran\textregistered~vessel
made of borosilicate glass of 250 ml and 500 ml~\cite{duran}. The
truncated conical sonotrode that conveyed ultrasounds was made of
AISI grade 304 steel. His dimensions (length, long diameter, short
diameter) and the dimensions of the threaded stub by which it was
screwed on the booster-piezoelectric unit were and have to be
designed in order to match the frequency of the mechanical
oscillations and reduce as much as possible any reflected power,
i.e. in order to have the maximum ultrasonic power transfer. This
adaptive design of the sonotrode is not unique but it is something
which has to be done case by case and strongly depends on the
material that the sonotrode is made of. Of course, once the long and
short diameters of the truncated cone are fixed the length of
sonotrode cannot just be determined by matching the resonance
condition, but there is a further constraint to be taken into
account. This constraint is the immersion of the sonotrode in the
solution where ultrasounds have to conveyed which has also to allow
for the diameter of the circular aperture of the vessel. We designed
its length in order to have a maximum immersion depth of about 4 cm
and a corresponding distance between the sonotrode tip and the
bottom of the vessel of about 5 cm.

All these geometrical dimensions are crucial to the positive
outcomes of the experiments as it will be clear further on. In all
the experiments, the cavitated solutions were made of deionized and
bidistilled water (18.2 M$\Omega$). Measurements of ionizing
($\alpha$, $\beta$ and $\gamma$) radiation background were carried
out, along with measurements of neutron radiation background. We
used three types of detectors of ionizing radiation: geiger counter
Gamma Scout~\cite{scout}  with a mica window transparent to
$\alpha$, $\beta$ and $\gamma$ radiation, and provided with two
aluminium filters 1 mm and 3 mm thick, to screen $\alpha$ radiation
and $\alpha$ and $\beta$, respectively; polycarbonate plate
detectors PDAC CR39 sensitive to ionizing radiation in the energy
range 40 keV -4 MeV and Tallium (Tl) activated, Sodium Iodine (NaI),
$\gamma$-ray spectrometer GAMMA 8000~\cite{amptek}.

The radiations $\alpha$, $\beta$ and $\gamma$, measured in all the
cavitation runs, turned out to be compatible with the background
radiation\footnote{This agrees with the results on the absence of
radiation emission in the first cavitation
experiments~\cite{cm1,cm2}.} \\
A magnetometer was used in order to take under control the local
magnetic field (always found compatible with the local magnetic
field of Earth, measured in absence of cavitation) and along with it
possible currents generated by the converting piezoelectric units
that might have affected the electronics of the geiger counters and
of the gamma spectrometer. Besides, in order to avoid any possible
interference through the power supplying wires and any possible
spurious communication among the electronic detectors through the
ground wire, the only electronic equipment to be connected to the
power network was the 20 KHz oscillation generator while all the
detectors were battery supplied. \\
Let's now focus our attention on the technique used to reveal the
possible neutron emission that we may expect during cavitation
according to the results of our previous
experiments~\cite{cm1,cm2,cm3}. The only hint that we got from these
experiments is the non conservation of the number of neutrons
according to the mass spectrometer analyses. In other words, the
only thing that we could expect was a possible neutron emission but
absolutely nothing could be said about its spectrum, its isotropy
and homogeneity in space and its constancy in time which could be
the most variable in terms of energy space and time. This wide range
of possibilities convinced us that the first step to be moved in
order to ascertain this hint was just to reveal the presence of
neutrons in a sort of a 'yes or no' detecting procedure and leave a
complete and more exhausting proper measurement to a second higher
and more accurate level of investigation grounded on the possible
positive answer from this first level of inquiry. Thus, we made our
choice and decided to use neutron passive detectors which are
capable of integrating neutron radiation within their energy range
regardless of the time feature of their emission. The passive
detectors that we used are called Defenders and are produced by BTI
(Bubble Technology Industries)\footnote{Let us notice that they are
no longer in production and have been replaced by similar devices.
However on the BTI there still is a web page dedicated to
them~\cite{def}}.

\begin{figure}
\begin{center}
\includegraphics{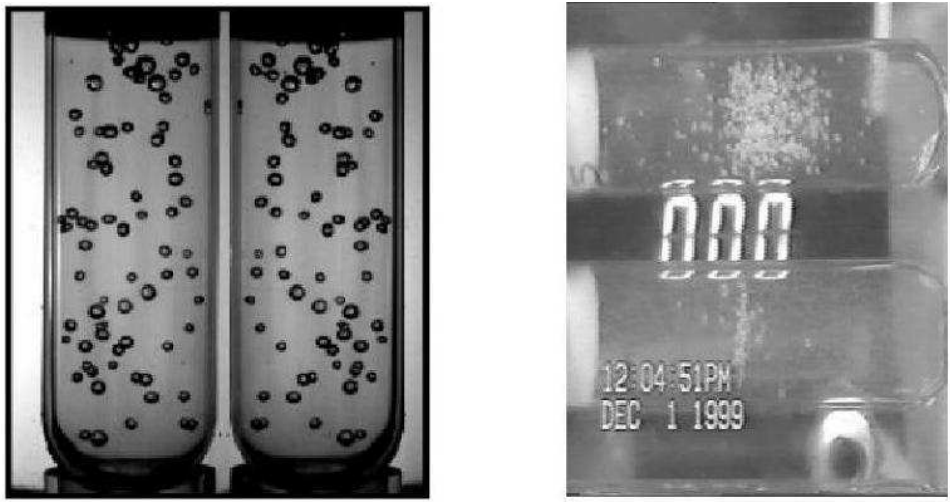}
\caption{Morphology and distribution of bubbles produced in a
Defender by the passage of neutrons (left); heavy ions (right). In
the second picture, the ion beam goes from bottom to
top.}\label{neutr-ion}
\end{center}
\end{figure}

   They consist of minute droplets of a superheated
liquid dispersed throughout an elastic polymer gel. When neutrons
strike these droplets, they form small gas bubbles that remain fixed
in the polymer. The number of bubbles is directly related to the
amount and the energy of neutrons, so the obtained bubble pattern
provides an immediate visual record of the neutron dose\footnote{
Each Defender was provided with its own calibration number (number
of bubbles/mRem) by which it was possible to convert the number of
bubbles into dose equivalent.}, see Fig.\ref{neutr-ion}.
We will be presenting two sets of experiments during which two
different kind of bubble detectors were used: Defender and Defender
XL. Their specifications are slightly different and will be reported
later on within each section describing a specific set of
experiments. Before moving on to the presentation of the experiments
and their results, it is important to stress at this stage some
features of these passive detectors and state what was done in order
to keep them under control. By doing this, we will also show that
the whole of neutron measurements of the first and second
investigation can be read as a sequence of control experiments which
allowed us to crosscheck by each of them the accuracy of the
previous one. The appearance of bubbles in these detectors can be
brought about by different sources.  Since the droplets are in a
metastable state they can be affected by heat and mechanical
compressions, just like ultrasounds. As to the heat, the first thing
that has to be stressed is that these detectors are temperature
compensated and their correct operation is guaranteed in the range
from 15\textdegree C to 35\textdegree C. Besides, the laboratory (a
small room) temperature was kept constant at about 20\textdegree C
$\pm$ 1\textdegree C by a heat pump that could work in reverse mode
as well. Of course we monitored by an infrared thermometer the
temperature of the Defenders all over their body and with particular
care on the area nearer to the vessel that became warm during
cavitation. The temperature of this specific part never exceeded
26\textdegree C which is well within the working temperature
guaranteed by the manufacturer. By comparing the number of bubbles
that popped up during each of the experiments of the first and
second investigation, one can unmistakably state that they cannot be
brought about by heat since all of the temperature increases of the
solutions treated by ultrasounds in all of the experiments were
always compatible with each other within $\pm$5\textdegree C, while
the number of bubbles ranged from less than ten up to 70 depending
on the applied ultrasonic power and the concentration of the
solutions. Let's now say something about the second possible source
of bubbles, i.e. ultrasounds. The minute droplets contained inside
the polymer gel are turned into bubbles as they receive the correct
amount of energy. Of course this amount can be conveyed to them by
mechanical compressions just like ultrasounds. Despite that, as it
will be clearly shown by the outcomes presented in the description
of the experiments, ultrasounds cannot be considered the cause of
the bubbles since the number of bubbles ranged from zero up to 70
while the power of ultrasounds, the distance between the vessel and
the detector were always the same and being mechanical vibrations
the cause of the bubbles, their number should have always been
nearly constant.
\section{Experimental Results}
\subsection{First Investigation}
Two separate investigations have been carried out. In the first one,
we subjected to cavitation five solutions of pure water and four
different salts in H$_2$O:

\begin{itemize}
\item 250 ml of bidistilled deionised water;
\item 250 ml with a concentration of 1 ppm of Iron Chloride FeCl$_3$;
\item 250 ml with a concentration of 1 ppm of Aluminium Chloride AlCl$_3$;
\item 250 ml with a concentration of 1 ppm of Lithium Chloride LiCl;
\item 500 ml with a concentration of 1 ppm of Iron Nitrate Fe(NO$_3$)$_3$.
\end{itemize}

Each of the first four cavitations lasted 90 min, while the Iron
Nitrate solution was cavitated both for 120 minutes. The schematic
layout of the experimental equipment is shown in Fig.\ref{first
layout}.

\begin{figure}
\begin{center}
\includegraphics[width=0.8\textwidth]{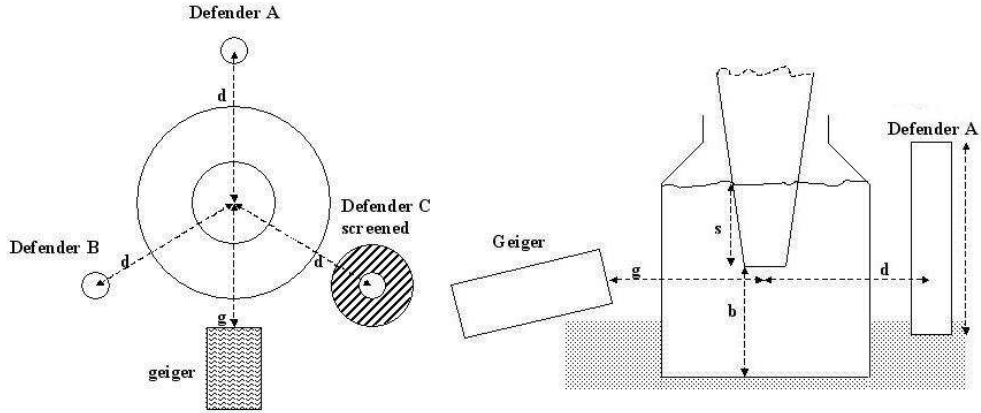}
\caption{Layout and lateral section of the experimental setup. d=7
cm, g=10 cm, s=4 cm, b=5 cm. This setup indicates that between the
cavitation area and the neutron detectors and the Geiger counter
there were 3.5 cm of water, the thickness of the Borosilicate (about
2 mm) and few centimetres of air.}\label{first layout}
\end{center}
\end{figure}

The cavitation chamber (vessel) was in the centre and the sonotrode
has to be imagined perpendicular to the plane of the figure, just
over the bottle and lined up with it. The immersion of the sonotrode
and the distance of its tip from the bottom of the vessel were about
4 cm and 5 cm respectively. For each cavitation experiment, we used
three neutron detectors Defender. They are cylinders 194 mm long
(their active part is 100 mm long) and with a diameter of 21 mm.
They are sensitive to neutrons in the energy range between 10 KeV
and 15 MeV. Their response is dose rate independent and their
minimum detection level is a tenth of an ounce of Plutonium in
seconds at 1 meter. Their response was determined to be about 100
counts/$\mu$Sv to $^{252}$Cf at 20\textdegree C. Their angular
response is isotropic and they are completely unaffected by gamma
radiation as it is stated by the manufacturer and it was
experimentally ascertained by irradiating them with a known source
of $^{60}$Co for several minutes without producing the tiniest
bubble. They were placed vertically and parallel to the vessel or
the sonotrode axis, arranged as shown in Fig.\ref{first layout}. One
of the Defenders was screened by immersing it in a cylinder of
carbon (moderator) 3 cm thick. The Geiger counter was pointed
towards the area inside the bottle where cavitation took place. A
second equal arrangement of three Defenders and the vessel
containing the same uncavitated solution (blank), was placed in a
different room and was used to measure the neutron radiation
background at the same time when cavitation was taking place. The
measurements of fast neutron radiation carried out in the
experiments with H$_2$O, Aluminium Chloride and Lithium Chloride
were compatible with the background level (20 nSv). On the contrary,
in the second and the fifth experiment, with Iron Chloride and Iron
Nitrate respectively, the measured neutron radiation was
incompatible with the neutron background level. The ultrasound power
and the experimental setup were the same for all of the five
experiments but only in two out of five we got a neutron signal
higher than the background. This evidence rules out ultrasounds as
the possible cause of bubbles in the Defenders. Neither could be the
heat generated by ultrasounds in the solutions since their
temperature and its rising time was always the same. Anyway, the
temperature of the body of the defenders never exceeded the starting
temperature (20\textdegree C) by more than 3\textdegree C being
perfectly within the range guaranteed and recommended by the
manufacturer. In the last thirty minutes of cavitation of the iron
salt solutions, the measured dose (\textasciitilde 100 nSv) was
significantly higher than (even 5 times) the background\footnote{The
neutron background measurements were carried out at the same time of
the cavitation, but in a different room, by means of equal detectors
placed around a similar vessel containing the same solution. The
results obtained were compatible with the background. The same
compatibility was found with detectors immersed in carbon both in
presence and in absence of cavitation. This last result is a further
confirmation of the neutronic origin of the bubble signals in the
Defenders.}

\begin{figure}
\begin{center}
\includegraphics{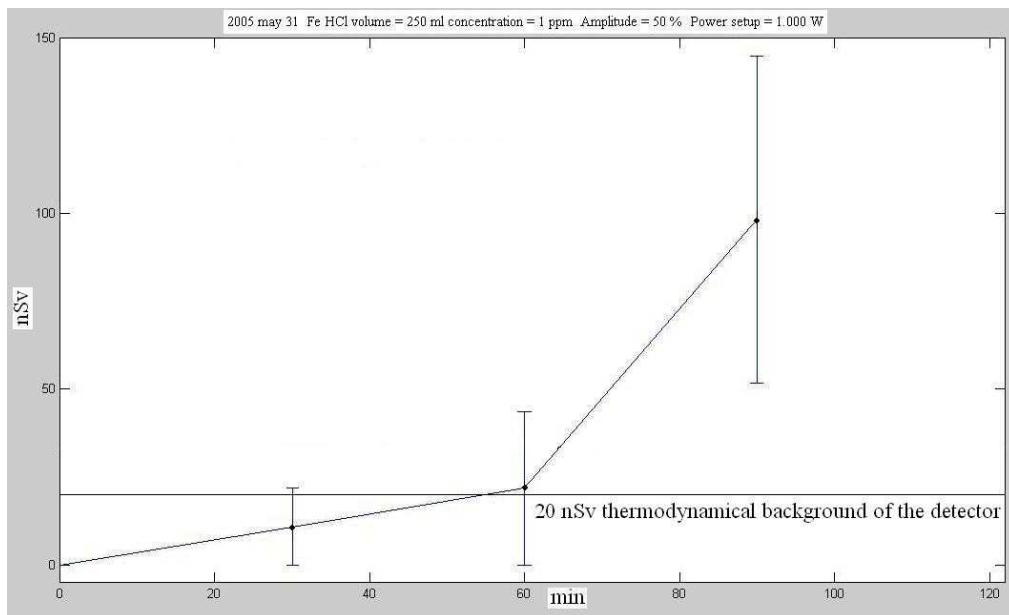}
\caption{Neutron dose (nSv) vs. cavitation time for Fe(Cl)$_3$
solution. The horizontal line represents the background level.
}\label{iron chloride graph}
\end{center}
\end{figure}

\begin{figure}
\begin{center}
\includegraphics{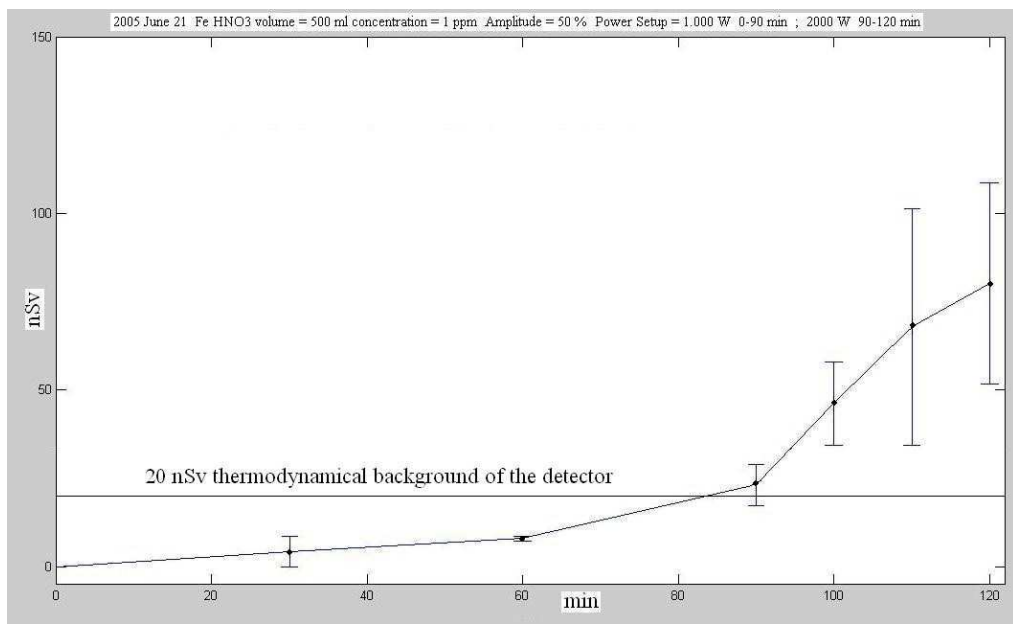}
\caption{Neutron dose (nSv) vs. cavitation time for Fe(NO$_3$)$_3$
solution. The horizontal line represents the background level.
}\label{iron nitrate graph}
\end{center}
\end{figure}

Precisely, the final measured dose was (98.50 $\pm$ 4.5) nSv for
FeCl$_3$ (Fig.\ref{iron chloride graph})and (76.00 $\pm$ 4.5) nSv
for Fe(NO$_3$)$_3$ (Fig.\ref{iron nitrate graph}).

The horizontal black line represents the sum of the measured
thermodynamical instability of the detectors\footnote{Indeed, when
the detectors are activated one faces an initial thermodynamical
instability due to the almost sudden decrease of pressure applied on
the superheated droplets dispersed in the gel. Some of them
evaporate and form bubbles which have to be taken into account as a
background level of blindness of the detector beyond a real,
although very low, neutron background level.} and of the measured
neutron background level and is equal to 20 nSv. In both graphs, the
values correspond to the mean of the two equivalent doses obtained
by the two defenders without moderator used during
cavitation\footnote{The number of bubbles was visually determined by
two of the experimenters independently and the mean value of the two
counts (which were always absolutely compatible and almost always
equal to each other) was taken as the number of bubbles to calculate
the dose.}. The error bars were determined by taking the root mean
square of the differences of the two equivalent doses and the mean
value. The increase of the derivative that appears quite evidently
in the last 30 minutes may be read as a first corroborating evidence
for the phenomenological considerations proposed in~\cite{cm4},
where two of us (F.C. and R.M.) proposed the existence of a
threshold in power and energy (and hence time) for piezonuclear
reactions to happen. In this sense, provided the ultrasonic power
transmitted into the solution is higher than the required
threshold~\cite{cm4}, the emission of neutrons produced by these
reactions begins only after that a certain amount of energy was
conveyed into the solution or, which is equivalently, after a
certain time interval. Let's now add a consideration which can be
drawn from reference~\cite{cm4} where the bubble collapse is
indicated as the main microscopical mechanism to induce piezonuclear
reactions and hence neutron radiation. The emission of neutrons does
not take place as from a stable source but, conversely, it happens
in bursts. This consideration can be considered at this stage as a
heuristic hypothesis which will be helpful in interpreting the
results of the second investigation, nevertheless some experimental
evidences presented further on will turn it into a sound empirical
hypothesis. The last fact of this first investigation was the
absence of ionizing radiation above the background level in all of
the experiments - even in those two in which we got the evidence of
neutron emission. Of course, this could mean either that gamma
radiation was not emitted at all as it usually is when neutrons are
emitted, or that the sensitivity of our detectors was not sufficient
to reveal their slight presence. Besides, we have to point out that
even if neutron emission took place without any consequent gamma
radiation\footnote{A possible explanation of this fact, based on a
space-time deformation of the interaction region between two nuclei,
can be found in ref.~\cite{cm4}.} from nuclei de-excitation, one
would expect gamma rays to be emitted from hydrogen capture anyway.
This first investigation permitted therefore to state that only the
presence of Iron in the cavitated solution gives rise to fast
neutron emission and therefore to nuclear processes induced by
cavitation.

\subsection{Second Investigation}

\begin{figure}
\begin{center} \
\includegraphics[width=0.8\textwidth]{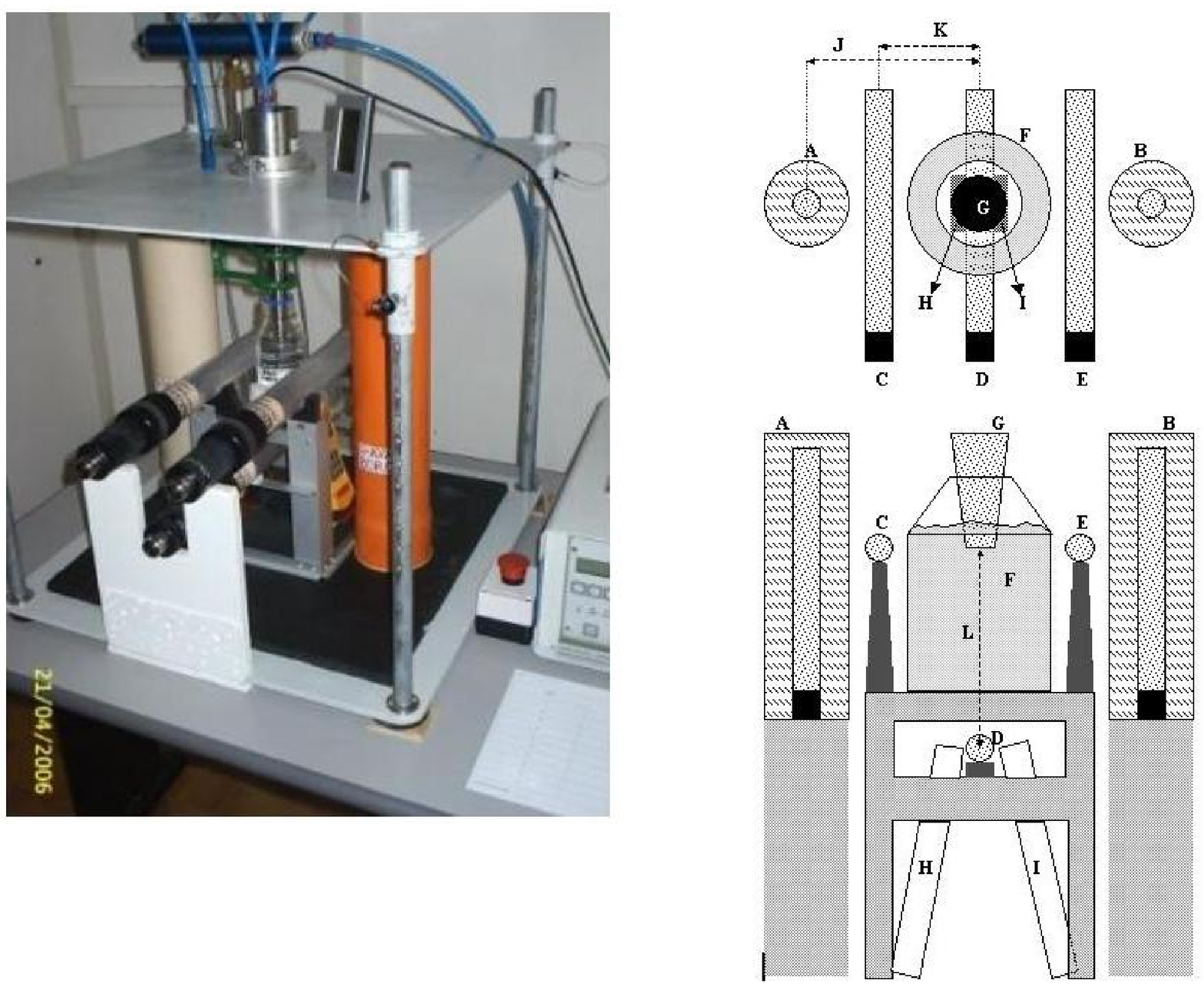} \caption{Experimental apparatus used
in the second investigation. The cavitation Chamber (F) is visible
in the middle of both pictures and the sonotrode, the vertical
tapered metal stick (G), is aligned with and inserted in it. The
green pipe surrounding the sonotrode conveyed the cooling air onto
the sonotrode surface. The three horizontal greyish cylinders
(C,D,E) with a black cylindrical endcap are the neutron detectors.
The two orange (right) (B) and creamy (left) (A) vertical cylinders
contained the two screened Defenders, one by boron (orange) and the
other by carbon (creamy). }\label{second layout}
\end{center}
\end{figure}

Since the first investigation highlighted the basic role of Iron in
producing piezonuclear reactions, the second one was devoted to a
systematic study of such an evidence, by using solutions with only
Iron Nitrate, since it gave rise, in the previous investigation, to
the maximum flux of emitted neutrons. Then, six cavitation runs
(each lasting 90 min) were carried out on the same quantity (250 ml)
of pure water and of a solution of Fe(NO$_3$)$_3$ with different
concentration, subjected to ultrasounds of different power. Namely,
the cavitated solutions could have three possible concentrations, 0
ppm (H$_2$O), 1 ppm and 10 ppm. Moreover, the oscillation amplitude
and hence the transmitted ultrasonic power took two different
values, 50\% and 70\%, corresponding to about 100 W and 130 W,
respectively. The energy delivered to the solution within the whole
cavitation time was 0.54 MJ and 0.70 MJ in the two cases. In order
to measure neutron radiation we employed five neutron detectors of
the Defender XL type, with higher sensitivity (by one order of
magnitude) with respect to those used in the first investigation.
These detectors are cylinders 47 cm long (their active part is 30cm
long) with a diameter of 5.7 cm. Their energy range lies between 10
KeV and 15 MeV. Their response is dose rate independent and their
minimum detection level is a hundredth of an ounce of Plutonium in
seconds at 1 meter. Their response was determined to be about 1000
counts/$\mu$Sv to $^{252}$Cf at 20\textdegree C. Their angular
response is isotropic and they are completely unaffected by gamma
radiation as it is stated by the manufacturer and it was
experimentally ascertained by irradiating them with a known source
of $^{60}$Co for several minutes without producing the tiniest
bubble. Background neutron measurements were accomplished at the
beginning of the whole set of cavitations. During each cavitation we
carried out ionizing radiation measurements by two Geiger counters
Gamma Scout~\cite{scout} , one with no aluminum filter and the other
with a 3 mm filter, used simultaneously. One picture and a layout of
the experimental apparatus used in the six cavitation runs are shown
in Fig.\ref{second layout}.

\begin{figure}
\begin{center} \
\includegraphics{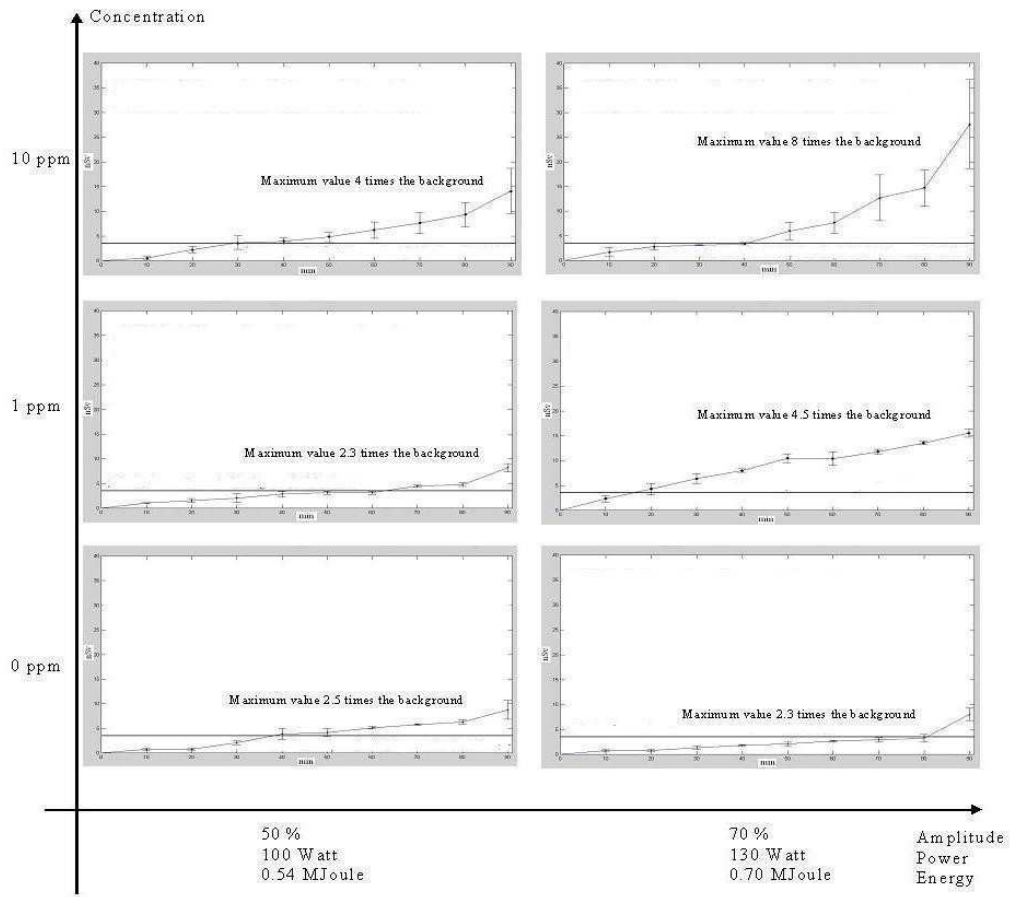} \caption{The six graphs (one for each cavitation of the second series) showing
the neutron dose (in nSv) as a function of time in minutes (time
interval 10 min). Each curve corresponds to one value of
concentration and one of the amplitude. The horizontal line in all
graphs corresponds to the thermodynamical noise of 3.5 nSv. The
graphs are displaced in a Cartesian plane, with concentration (in
ppm) on the y-axis and amplitude (power) on the x-axis.
}\label{graph graphs}
\end{center}
\end{figure}

 The vessel in which cavitation took place (F) (cavitation
chamber, the same as the first investigation) is visible in the
middle of both pictures and the sonotrode, the vertical tapered
metal stick, is aligned with and inserted in it. The three
horizontal greyish cylinders with a black cylindrical endcap are the
neutron detectors. Two of them (C) and (E) were positioned next to
the chamber at a height with respect to the tip of the sonotrode, in
order to be struck by horizontally emitted neutrons. Their distance
(K) from the centre of the vessel is half the diameter of the bottle
(3.5 cm of water) plus the thickness of the borosilicate glass 2 mm
and 5 mm of air. The third detector (D) was placed underneath the
chamber in order to collect the vertically emitted neutrons. Since
in this second investigation we reduced the immersion of the
sonotrode (G) to 1 cm, the distance (L) of the Detector (D) from the
sonotrode tip is in this case the sum of 9 cm of water, 2mm of
borosilicate glass, 4 mm of Plexiglas and 3 cm of air. The two
vertical cylinders (A and B) contained one neutron detector each, of
the same type of the three horizontal ones. The detectors were
surrounded, and hence screened, by 3 cm of Boron powder (B) (thermal
neutron absorber) and by 3 cm of Carbon powder (A) (neutron
moderator), respectively. The distance (J) of these two screened
Defenders XL from the axis of the vessel was the sum of the diameter
of the bottle (water) plus the thickness of the borosilicate glass
(2 mm), 10 cm of air, 1 mm of PVC and 3 cm of either Boron or
Carbon.Two geiger counters (H and I) were pointed towards the bottom
of the cavitation chamber, one with unscreened mica window, the
other with a shield of 3mm of Aluminium. The distance of the mica
window from the sonotrode tip was again (L) as specified above. In
all of the six experiments of this second investigation, the three
horizontal, unscreened Defender XL's measured a neutron emission
significantly higher than the background level. The two vertical,
screened Defender XL's (both by boron and carbon) always detected a
reduced neutron dose, comparable with the background one (thus again
providing further evidence of the neutron origin of the bubble
signals). For all of the six experiments, we plotted the measured
doses of neutrons (in nano-Sievert) as function of the cavitation
time. The number of bubbles was counted every 10 min. Each curve
corresponds to one concentration of the Fe(NO$_3$)$_3$ solution,
from 0 ppm to 10 ppm, and one oscillation amplitude (and therefore
ultrasonic power), 50\% (100 W) or 70\% (130 W). The six graphs are
reported in Fig.\ref{graph graphs}.

They are displaced in a Cartesian coordinate system with
concentration on the y-axis and amplitude (power) on the x-axis. As
in the first investigation, the horizontal black line represents the
sum of the measured thermodynamical instability of the detectors and
of the measured neutron background level. The examination of the six
graphs of Fig.\ref{graph graphs} does not report the threshold
behaviour in energy that we found in the first investigation, namely
the sharp and sudden increase of the curve derivative in the last 30
minutes of cavitation. Despite that, according to our heuristic
hypothesis (which will be experimentally supported in the following)
about the neutron emission taking place in bursts, it will be wrong
to interpret these curves as a sign of a stable neutron emission.
Conversely, still considering valid this hypothesis, one can ascribe
this different behaviour between the first and the second
investigations to the different immersions of the sonotrode in the
solution, which was about 4 cm (about 5 cm from the bottom of the
vessel) in the former and only 1 cm (about 10 cm from the bottom of
the vessel) in the latter. This means that both the height of the
neutron peaks (bursts) and, hence, the emitted dose can be
controlled somehow by this geometrical parameter. This consideration
allows one to ascribe this apparent lack of threshold behaviour to
the reduced height of the neutron peaks emitted during the
cavitations performed in the second investigation with respect to
those emitted in the first one. This reduced height spread the
neutron dose over a longer period of time preventing the threshold
behaviour from showing up. It will be the purpose of our future
investigations to establish the time of appearance of the first
neutron burst and verify whether it takes place beyond the energy
(or time) threshold. Moreover, Fig.\ref{graph graphs} further
disproves the possible criticism about a possible generation of the
bubbles by ultrasounds rather than by neutrons. Indeed, by looking
at the compound graph and reading it along its columns, i.e. keeping
the amplitude (power) fixed, it is seen that the curves are
different, while the ultrasonic power is always the same.
Conversely, had ultrasounds been the real cause of the bubbles, one
should have had equal effects. Besides, we add that the temperature
of the laboratory was stabilized to 20\textdegree C by a heat pump,
which could work in reverse mode as well. Moreover, we checked every
ten minutes the temperature of the body of the two defenders XL next
to the cavitation chamber and in particular of that part close to
the warm vessel. The temperature of this part increased gradually
from 20\textdegree C but never exceeded 25\textdegree C which is
perfectly within the working range (15\textdegree - 35\textdegree C
) guaranteed by the manufacturer who thermally stabilized their
operation. As a further proof against any possible influence of
temperature or IR irradiation on the number of bubbles in the
defenders, we checked that at equal temperature of the solution in
the vessel, and equal ultrasonic power, the bubble distribution in
the defender XL did not show any systematic concentrations
(qualitatively and quantitatively in term of number of bubbles) near
the warmest part of the vessel and in the surroundings where
possible thermal gradients might have had some effect on the
stability of the defenders. Let us also remark that in the second
investigation one got evidence for neutron emission also in
cavitating pure water, unlike the case of the first one. This is
obviously due to the higher sensitivity of the detectors employed in
the second investigation. Such a result agrees with the indirect
evidence for neutron emission obtained in the first experiment of
water cavitation, in which the changes in concentration of the
stable elements occurred with a variation in neutron
number~\cite{cm1,cm2}. At the light of the above results, we can say
that the cavitating device behaves as an ultrasonic nuclear reactor.
As we have already said, we performed measurements of the ionizing
radiation by means of the above mentioned (filtered and unfiltered)
Geiger counters. The measured radiation was always compatible with
the background level. As a further check of the absence of $\gamma$
radiation, we carried out, in absence of cavitation and during
cavitation of Iron Nitrate (70\% amplitude, concentration $>$10 ppm,
duration 90 mins), simultaneous measurements by means of the two
Geigers and through a tallium (Tl) activated, Sodium Iodine (NaI),
$\gamma$-ray spectrometer. We found again a perfect compatibility
between the background spectrum and that during cavitation both for
the two Geigers and for the NaI (Tl), $\gamma$-ray spectrometer (in
spite of the neutron signal with maximum of (9.1 $\pm$ 0.5) nSv
measured by the Defender XL's). Thus, the results of the second
investigation too provided evidence for the emission of anomalous
nuclear radiation, since neutrons were not accompanied by gamma
rays. These outcomes about the apparent absence of gamma rays have
to be commented by what we have already said above for the first
investigation. The NaI(Tl)  spectrometer allowed us to increase by
several orders of magnitude the accuracy and sensitivity of gamma
ray detection. Despite that, we need again to raise the question
about the lack of gamma rays from Hydrogen capture which will have
to be addressed to the future experiments.

The systematic analysis carried out by cavitating water solutions of
Iron Nitrate, for all of which evidence of neutron radiation was
gotten, shows that the phenomenon is perfectly reproducible.
Moreover, we have been able, by changing the immersion depth of the
sonotrode tip, to reduce the emitted neutron dose by one order of
magnitude. In fact, in the last cavitation run we got a maximum of
(28.0 $\pm$ 7) nSv. This implies that the phenomenon can be somehow
controlled.

\subsection{Further check and features of neutron emission}
\begin{figure}
\begin{center} \
\includegraphics[width=0.8\textwidth]{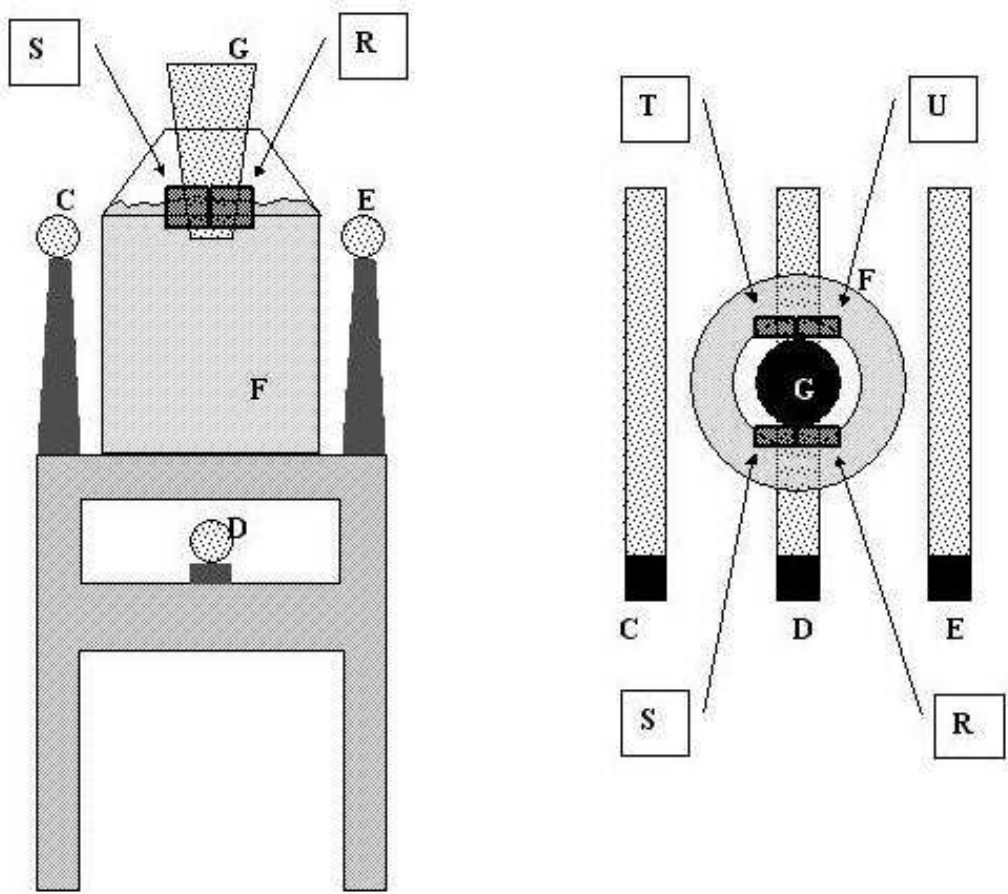} \caption{Layout of the experimental set-up of the second investigation showing the position
of the Boron screened CR39 plates with respect of the rest of the
equipment.}\label{cr39}
\end{center}
\end{figure}

In the previous two investigations, the evidence for neutron
emission was highlighted by means of the detectors Defender through
the analysis of the bubble signals. As a further check, we carried
out a further experiment utilizing not only the Defender XL's but
also boron-screened CR39 detectors according to a well known
technique~\cite{tommasino,khan,ize}. By the same experimental
apparatus used in the second investigation (see Fig.\ref{second
layout}), we subjected to cavitation 250 ml of a water solution of
Iron Chloride (FeCl$_3$) with concentration 10 ppm. The cavitation
lasted 90 min at the ultrasound frequency of 20 KHz, with
oscillation amplitude of 70\% of the maximum amplitude,
corresponding to a power of 130 W (namely to a total energy of 0.70
MJ). The choice to use again a solution of FeCl$_3$ was due to the
fact that, all the other conditions being equal, we noted that with
Iron Chloride there is a higher release of macroscopic energy than
with Iron Nitrate (the liquid evaporation is from 2 to 5 times that
observed with the latter solution). Due to the equality of
thermodynamical conditions, this cannot be explained in terms of
ultrasounds only. The two unscreened lateral Defender XL's (C and E)
measured a maximum dose of neutrons of 14.5 nSv, 4 times higher than
the detector thermodynamic noise of 3.5 nSv. Moreover, we placed,
externally to the cavitation chamber, two pairs of 1 cm by 1 cm
plate CR39 detectors (R,S and T,U) as shown in Fig.\ref{cr39}.

\begin{figure}
\begin{center} \
\includegraphics[width=0.8\textwidth]{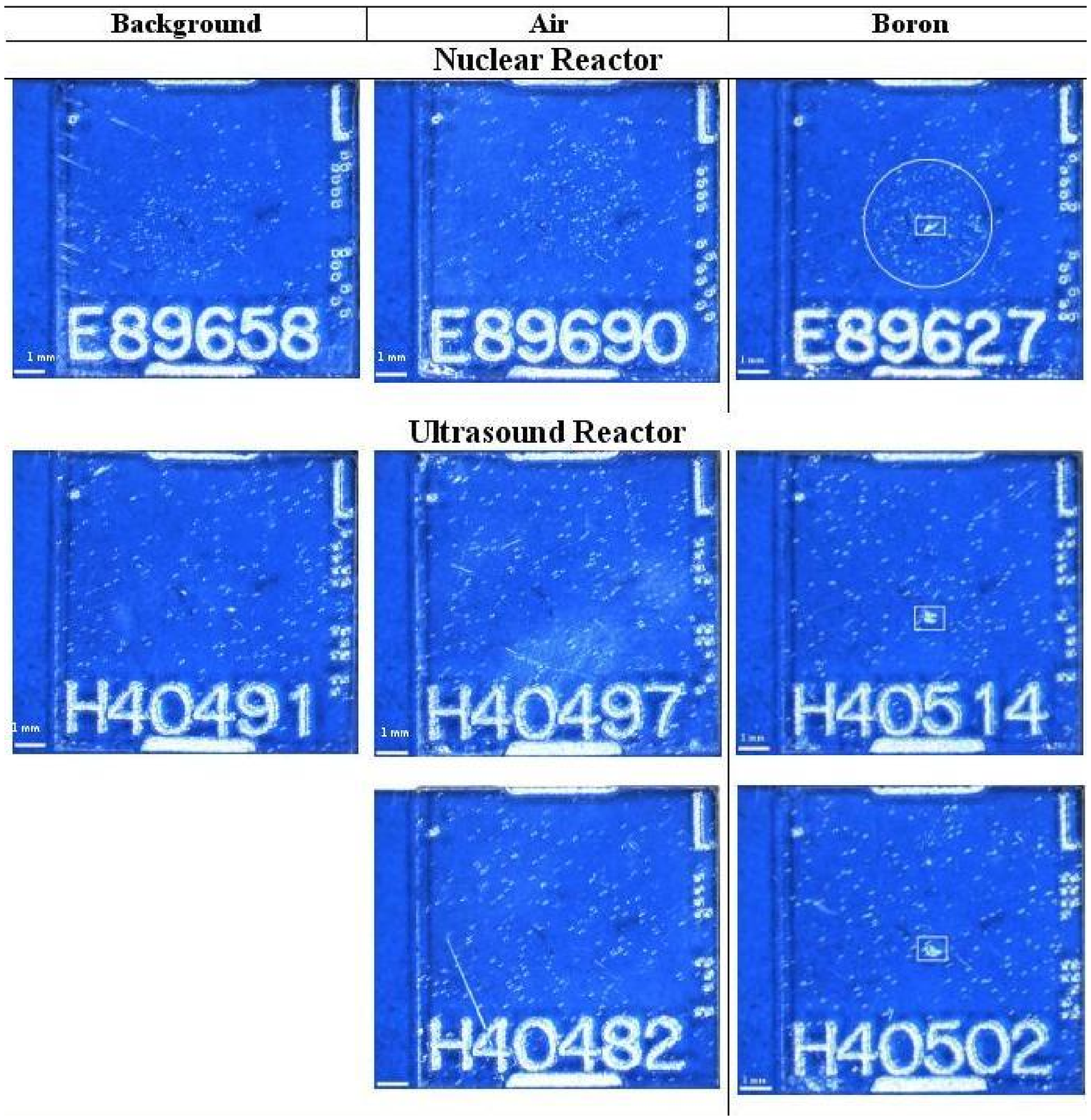} \caption{Showing the traces left by neutrons on the CR39 detecting plates for
the two cases of the nuclear reactor TAPIRO and the ultrasonic
reactor. The magnification is 10X. The three columns from left to
right refer, respectively, to the background, the CR39 in air and
the CR39 immersed in boron. In the third column, the rectangles
enclose the traces of the maximum neutron intensity (corresponding
to the beam axis in the case of the nuclear reactor).
}\label{cr39trace}
\end{center}
\end{figure}

Each plate was at a distance of about 4 cm from the vertical axis of
the cavitation chamber, at the same level of the sonotrode tip. In
between the CR39 plates and the axis of the vessel there were 3.5 cm
of the solution, 2 mm of the borosilicate glass and about either 3
mm of air or 3 mm of Boron. The two couples were diametrically
opposite to each other. In each pair, a CR39 was in air (S and T),
whereas the other detector was immersed in boron (R and U) (whose
interaction with neutrons gives rise to alpha radiation to which
CR39 are sensitive). The results obtained are displayed in the
second and third row of Fig.\ref{cr39trace}. By the boron CR39 we
were able to detect neutrons with energies below 10 KeV too and,
above all, thermal neutrons.

In order to have an idea of what the traces should look like on
these detectors after etching, four more detectors were irradiated
by neutrons using as source, the fast neutron nuclear reactor TAPIRO
at Casaccia ENEA Rome, the neutron equivalent dose conveyed onto the
detectors was 2.1 $\mu$Sv through a diagnostic neutron
channel\footnote{Not knowing what kind of neutron spectrum to expect
from the cavitated solution, as already stated, we decided to
produce our comparison model of traces by a source whose spectrum
were the widest possible, i.e. a nuclear reactor. According
to~\cite{tommasino} these kind of detectors can detect fast,
epithermal and thermal neutrons with different sensitivities of
course. Hence the integral effect on the detectors, due to almost
the whole neutron spectrum, would be traces whose quantity and shape
would be compared to those obtained from the piezonuclear reactor.
As already stated, the main target of these investigations is to
reveal the presence of neutrons in a sort of a 'yes or no' detecting
procedure. In future investigations we will perform more
quantitatively accurate measurements by calibrating the CR39
detectors by known neutron energy sources.}. The output channel of
TAPIRO was calibrated to get a neutron equivalent dose rate of 21
$\mu$Sv/h. A boron CR39 was used to measure the background level
around the reactor, other two, one in air and the other immersed in
boron, were placed at about 3 m from the reactor core and radiated
for 5 min. The results are shown in the first row of
Fig.\ref{cr39trace}. The comparison between the traces produced by
neutrons in the CR39 immersed in boron (third column) in the nuclear
reactor case (first row) and in the ultrasound one (second and third
row) shows that their pattern (although not their extension) is
perfectly similar. It is also possible to notice that the area of
the thick trace produced by the reactor neutrons is about half of
the areas of the thick traces produced by the neutrons generated by
ultrasounds during cavitation. The Boron CR39 detectors can reveal
neutrons of any energy. While fast neutrons are not affected by
Boron and leave their own traces on the polycarbonate surface, slow
neutrons and thermal neutrons, above all, convert into alpha
particles by interacting with Boron-10 ($^{10}$B) (according to
$^{10}$B(n,$\alpha$)$^7$Li) and through this mechanism produce a
much wider and deeper trace on the polycarbonate surface than fast
neutrons. If we use this fact and we compare the CR39 traces
obtained in this experiment (compatible with equivalent doses of 4-5
$\mu$Sv in 90 minutes), with the bubble signals collected by the
Defender XL's in this same experiment (14.5 nSv in 90 minutes), and
with those by the Defenders of the first Investigation (between 80
and 100 nSv in 90 minutes), we are allowed to conclude that the bulk
of the neutron emission corresponds to neutrons having energy in the
low epithermal range and even lower. We believe that the outcomes
shown by these photos represent a fairly sound proof to corroborate
our heuristic hypothesis about the emission of neutrons in bursts.
The trace pattern together with the thick trace on the CR39 plate
(like E89627 in Fig.\ref{cr39trace}), that was in front of the
nuclear reactor, suggests that the emission of neutrons from the
reactor core is constant and isotropic. Of course, the reactor
channel acted as a filter which selected those neutrons whose
velocity was almost parallel to the channel axis. These neutrons
produced the thick track right on the channel axis and that almost
circular distribution highlighted on the plate E89627, but somehow
visible on the plate E89690 too. These effects were collected within
5 minutes. On the contrary, despite the cylindrical symmetry of our
experimental equipment (the vessel and the sonotrode), it is fairly
clear that the neutron emission during cavitation was neither
constant nor isotropic. Were it isotropic, one would have got a more
uniform distribution of traces and more thick traces on the CR39
plates and a more uniform distribution of bubbles in the defenders.
As to the constancy of emission, one would face the fact that the
microscopical mechanism that brings about neutron emission is bubble
collapse, which is governed by quite a few variables, like bubble
dimension, quantity and type of atoms on the bubble surface. All
these variables, completely uncontrolled yet, make neutron emission
more likely an impulsed process rather than constant. In this sense,
neutron emission takes place in bursts at different instants of
time, along diverse space directions and with different height and
energy spectrum.

\section{Coherence with the findings of other experiments}
Our cavitation experiments performed in the last decade evidenced
two kinds of phenomena: production of nuclides
(experiments~\cite{cm1,cm2,cm3}) and neutron emission (present
experiments). Let us discuss such findings in connection with the
results of other experiments. As to nuclide production, the findings
of the previous experiments (in particular of the first
one~\cite{cm1,cm2}) are similar under many respects to those
obtained by Russian teams at Kurchatov Institute and at Dubna
JINR~\cite{uru1,uru2,kuz,volk,uru3} in the experimental study of
electric explosion of titanium foils in liquids. In a first
experiment carried out in water, the Kurchatov
group~\cite{uru1,uru2} observed change in concentrations of chemical
elements and the absence of significant radioactivity. These results
have been subsequently confirmed at Dubna~\cite{kuz}. Recently, the
experiments have been carried out in a solution of uranyl sulfate in
distilled water, unambiguously showing~\cite{volk} a distortion of
the initial isotopic relationship of uranium and a violation of the
secular equilibrium of $^{234}$Th. Due to the similarity of such
results with ours, in our opinion the two observed phenomena have a
common origin. Namely, one might argue that the shock waves caused
by the foil explosion in liquids act on the matter in a way similar
to ultrasounds in cavitation. In other words, the results of the
Russian teams support the evidence for piezonuclear reactions.
However, let us notice that this is by no means a completely new
result. Indeed, we recall that in the past some
investigations~\cite{dieb,kali,kozi,wint} have highlighted the
ability of pressure and shock waves to generate autocatalytic
fission-fusion reactions in compounds containing also uranium,
tritium and deuterium. In such experiments, neutron fluxes have been
observed in the range 10$^7$ -10$^{13}$ neutrons/cm$^2$s. As to
neutron emission, we already quoted the Oak Ridge
experiment~\cite{tal1,tal2,tal3,tal4} on possible nuclear fusion in
deuterated acetone subjected to cavitation. The measured neutron
flux was said to be compatible with d-d fusion during bubble
collapse. Some authors disclaimed the results~\cite{ss}, others
conversely confirmed them~\cite{xb,for}. As to what the results of
our investigations are, one would not be surprised of the
controversial results and hence opinions on the outcomes of the Oak
Ridge experiments~\cite{tal1,tal2,tal3,tal4}. Our outcomes show that
neutron emission is obtained by cavitating solutions containing Iron
and, even if in a very small quantity, by cavitating  pure water.
Hence the effects, that we measured, must be brought about by almost
thoroughly unknown mechanisms which are triggered by pressure. With
this in mind, we believe that whoever tried to reproduce the Oak
Ridge experiments must have faced unusual behaviours and results
since along with the very well known and expected neutrons from D-D
fusion, other unknown effects (like the existance of a time (energy)
threshold for neutron emission) would be superimposed, and would
generate confused results which do not precisely confirm the common
phenomenological predictions about fusion.\\The
experiments~\cite{tal1,tal2,tal3,tal4} belong to the research stream
known as sonofusion (or acoustic inertial confinement fusion),
pioneered by Flynn in 1982~\cite{bren}. It amounts to the attempt to
produce known nuclear reactions by means of ultrasounds and
cavitation. Conversely our case is completely different. We produced
new nuclear reactions (piezonuclear reactions) that involve heavy
nuclei but do not, apparently, affect Hydrogen or light ones (at
least within 90 minutes) under unusual conditions like the existence
of an energy threshold for these reactions to happen and like the
apparent lack of gamma emission concomitant to neutron emission
(although this needs to be confirmed).

\section{Conclusion}
The experiments we carried out permit therefore to conclude that the
cavitation process is able to induce in Iron salt solutions emission
of either fast and epithermal neutrons. This constitutes a further
evidence for piezonuclear reactions. Moreover, we have been able to
state some fundamental features of such a neutron emission, namely:
1) it exhibits threshold behavior in power, energy and time; 2) it
occurs in anomalous conditions, namely without concomitant sensible
production of $\gamma$-rays. If independently confirmed, our results
would probably constitute a signature of new physics.\\Let us
conclude by putting forward a conjecture  about these piezonuclear
reactions and foretell that they can be brought about by properly
compressing solid materials that contain iron (e.g. granite), for
instance in one of those toughness experiments that are very common
in Mechanical and Civil Engineerings. More precisely, it will be
possible to measure neutron emission at the instant of fracture of
the specimens of these materials as their compression increases and
reaches the breaking load. According to what is being done for
liquids, it will be necessary to study neutron emissions as function
of the compression speed of the specimens.
\\

\textbf{Acknowledgments}. We are greatly indebted to all people who
supported us in many ways in carrying out the experiments: the
military technicians of the Italian Armed Forces A. Aracu, A.
Bellitto, F. Contalbo, P. Muraglia; M. T. Topi, Director of ARPA
Laboratories of Viterbo; the following personnel of Casaccia ENEA
Laboratories: P. Giampietro, Director, G. Rosi, responsible of the
nuclear reactor "TAPIRO", and A. Santagata; G. Ingo and C. Ricucci,
of the Microscopy Laboratory ISMN-CNR and L. Petrilli, of CNR-Rome 1
Area, Montelibretti, for performing the mass spectrometry of the
cavitated samples; R. Capotosto, Department of Physics "E. Amaldi"
of University "Roma Tre", for technical support on the sonotrode
tip. On the theoretical side, invaluable comments by E. Pessa are
gratefully acknowledged. Thanks are also due to F. Mazzuca,
President of Ansaldo Nucleare, for deep interest and warm
encouragement. Last but not least, it is both a duty and a pleasure
to thank in a special way Fabio Pistella, former President of CNR
(Italian National Research Council) for his caring assistance,
continuous interest, convinced participation to all stages of the
experiments, and eventually for his experience of nuclear physicist
he kindly provided to us during the discussions concerning the
experimental results.

\end{document}